\documentclass[fleqn,usenatbib]{mnras}

\usepackage{newtxtext,newtxmath}

\usepackage[T1]{fontenc}

\DeclareRobustCommand{\VAN}[3]{#2}
\let\VANthebibliography\thebibliography
\def\thebibliography{\DeclareRobustCommand{\VAN}[3]{##3}\VANthebibliography}

\usepackage{graphicx}	
\usepackage{amsmath}	
\usepackage{xcolor}
\usepackage{footnote}
\usepackage{pdflscape}
\usepackage{url}

\DeclareUnicodeCharacter{02BC}{\ensuremath{'}}

\title[]{An Escaping Outflow in a  Galaxy with an Intermediate-mass Black Hole}

\author[Zheng et al.]{
Zhiyuan Zheng,$^{1}$
Yong Shi,$^{1,2}$\thanks{E-mail: yong@nju.edu.cn}
Fuyan Bian, $^{3}$
Xiaoling Yu,$^{4}$
Junfeng Wang, $^{5}$
Jianhang Chen, $^{6}$
\newauthor
Xin Li, $^{1}$
and Qiusheng Gu, $^{1,2}$
\\
$^{1}$School of Astronomy and Space Science, Nanjing University, Nanjing      210093, Peopleʼs Republic of China\\
$^{2}$Key Laboratory of Modern Astronomy and Astrophysics (Nanjing         University), Ministry of Education, Nanjing 210093, People’s Republic of China\\
$^{3}$European Southern Observatory, Alonso de C´ordova 3107, Casilla 19001, Vitacura, Santiago 19, Chile\\
$^{4}$College of Physics and Electronic Engineering, Qujing Normal University, Qujing 655011, P.R. China\\
$^{5}$Department of Astronomy, Xiamen University, Xiamen, Fujian 361005, China\\
$^{6}$European Southern Observatory (ESO), Karl-Schwarzschild-Strasse 2, D-85748 Garching, Germany\\
}

\date{Accepted XXX. Received YYY; in original form ZZZ}

\pubyear{2023}

\begin{document}
\label{firstpage}
\pagerange{\pageref{firstpage}--\pageref{lastpage}}
\maketitle

\begin{abstract}

While in massive galaxies active galactic nuclei (AGN) feedback plays an important role, the role of AGN feedback is still under debate in dwarf galaxies. With well spatially resolved data obtained from the Multi-Unit Spectroscopic Explorer (MUSE), we identify a spatially extended ($\rm \sim 3\; kpc$) and fast ($V_{80} \sim 471\; \rm km\;s^{-1}$) AGN-driven outflow in a dwarf galaxy: SDSS J022849.51-090153.8 with $M_{*} \sim 10^{9.6}\;{\rm M_{\odot}}$ that host an intermediate-mass black hole of $M_{\rm BH} \sim 10^5\;{\rm M_{\odot}}$ and $L_{\rm AGN}/L_{\rm Edd} \sim 0.15$. Through the measurement of the rotation curve, we estimate the escape velocity of the halo and the ratio of the outflow velocity to the halo escape velocity to be $1.09\pm0.04$, indicating that the outflow is capable of escaping not only the galaxy disk but the halo. The outflow size of our AGN is found to be larger than AGN in massive galaxies at the given AGN [\ion{O}{iii}] luminosity, while the size of the photo-ionized narrow-line region is comparable. These results suggest the important role of AGN feedback through outflows in dwarf galaxies when their central intermediate-mass black holes accrete at high-Eddington ratios.

\end{abstract}




\section{Introduction}

Observations have shown that there are correlations between the physical properties of supermassive black holes (SMBH) and their host galaxies \citep{2013ARA&A..51..511K}. By accreting the surrounding gas material, SMBHs shine as the active galactic nuclei (AGN), meanwhile, their feedback processes affect the evolution of host galaxies \citep{2015ARA&A..53..115K}. In the theoretical framework of galaxy formation and evolution, such feedback brings simulated galaxy properties in agreement with observations
by suppressing star formation in massive galaxies
\citep{2014MNRAS.444.1518V,2016MNRAS.463.3948D,2018MNRAS.475..648P, 2013MNRAS.428.2966P, 2015MNRAS.450.1349K, 2015MNRAS.446..521S}.

The AGN-driven outflow is proposed to be an effective way to affect the physical properties of host galaxies \citep{2015ARA&A..53..115K}. Observational evidence has shown that the presence of AGN-driven outflow in massive galaxies represents an ongoing AGN feedback process \citep{2011ApJ...729L..27R,2013ApJ...775L..15R,2015ApJ...801..126R,2013MNRAS.430.2327L,2013MNRAS.436.2576L,2014MNRAS.441.3306H,2017ApJ...850...40R,2019MNRAS.487L..18R}. Through such powerful and spatially extended outflow, AGN feedback is sufficient to affect the baryonic matter of galaxies out to their circumgalactic media (CGM) \citep{2014ApJ...790..116V,2018ApJ...857..126L}.

In dwarf galaxies, a common picture holds that stellar feedback is the primary feedback mechanism \citep{2017arXiv170109062H, 2018ApJ...855L..20M, 2021MNRAS.507.2423S}, while the role of AGN feedback is still under debate. New simulations from \citet{2019MNRAS.484.2047K,2021MNRAS.503.3568K} pointed out that AGN may play an indirect role in regulating the baryonic matter cycle in dwarf galaxies. Meanwhile, other simulations also rise the possibility of direct AGN feedback in dwarf galaxies via gas ejection \citep{2018MNRAS.473.5698D,2019MNRAS.487.5549B, 2022MNRAS.516.2112K, 2022arXiv221105275S}. With the development of observational techniques, ones have detected more and more AGN activities in dwarf galaxies \citep{2008ApJ...688..794S,2016ApJ...829...57B, 2018ApJ...863....1C,2021ApJ...911...70B,2022ApJ...937....7S}.  \citet{2018MNRAS.476..979P} presented that the AGN feedback may affect the quenching process of dwarf elliptical galaxies in the MaNGA survey. \citet{2019ApJ...884...54M} observed 29 dwarf galaxies with AGN through the Keck LRIS long-slit spectroscopy.  Nine of their 29 objects show spatial extended and fast AGN-driven outflows, which suggests that AGN feedback may play an important role in their dwarf galaxy sample. \citet{2020ApJ...905..166L} selected eight AGN in dwarf host galaxies from \citet{2019ApJ...884...54M} and drew similar conclusions through integral field spectroscopy (IFS) observations with Keck KCWI and Gemini GMOS. 

By assuming a simple density profile, such as an isothermal sphere model, the escape velocity is often derived from the circular velocity, i.e., $V_{\rm esc} \approx (2.6-3.3)\ V_{\rm circ}$ \citep{2020A&ARv..28....2V}. In massive galaxies, cool outflows (atomic and molecular outflow) are hard to escape from the halo \citep{2015ApJ...804...83S, 2016ApJ...833...39S, 2022MNRAS.516..861T}. The situation does not become better in warmer material, since the outflow velocity of ionized gas is not always larger than the neutral gas \citep{2017ApJ...850...40R, 2020A&ARv..28....2V}.

However, the situation in dwarf galaxies may be quite different. In some objects where the ionized gas outflow is detected, the ratio of outflow velocity and escape velocity (i.e. $ V_{\rm out}/V_{\rm esc}$) decreases with the galaxy mass \citep{2014A&A...568A..14A, 2019MNRAS.486..344R}, and drops to a unity around the stellar mass of $4\times10^{9}{\rm M_{\odot}}$ so that $ V_{\rm out}>V_{\rm esc}$ \citep{2019MNRAS.490.4368S}. These results indicate that the outflow escapes easier in dwarf galaxies, at least for the ionized gas. However, the escape velocity is difficult to measure in dwarf galaxies, usually based on indirect methods such as the abundance matching \citep{2019ApJ...884...54M}. The direct method with the rotation curve requires data reaching the flat part of the rotation curve, which is sometimes hard especially in dwarf galaxies where AGN-driven outflow is detected \citep{2020MNRAS.498.4562M, 2020ApJ...905..166L}. 

In this work, we carry out detailed studies of AGN in a dwarf galaxy SDSS J022849.51-090153.8 (SDSS J0228-0901) with Multi-Unit Spectroscopic Explorer (MUSE) on Very Large Telescope (VLT) to probe properties of its outflows. It hosts an intermediate-mass black holes with $M_{\rm BH}\sim3.7\times10^5{\rm M_{\odot}}$ \citep{2018ApJ...863....1C}. Benefiting from the large field of view of MUSE, we have a chance to measure the escape velocity more precisely. This paper is organized as follows. In \S~2, we introduce the basic properties of SDSS J0228-0901 and our observations. Then the data analysis is presented in \S~3. Our investigation results and understandings are described in \S~4. Finally, \S~5 gives a summary. We assume a $\rm \Lambda CDM$ cosmology of $H_0 = 67.4$   $\rm km\;s^{-1}\;Mpc^{-1}$, $\Omega_{\rm m} = 0.315$ and $\Omega_{\rm \Lambda} = 0.685 $ \citep{2020A&A...641A...6P}.

\section{Observation and Data Reduction}

\begin{figure*}
   \resizebox{\hsize}{!}
             {\includegraphics[width=\textwidth]{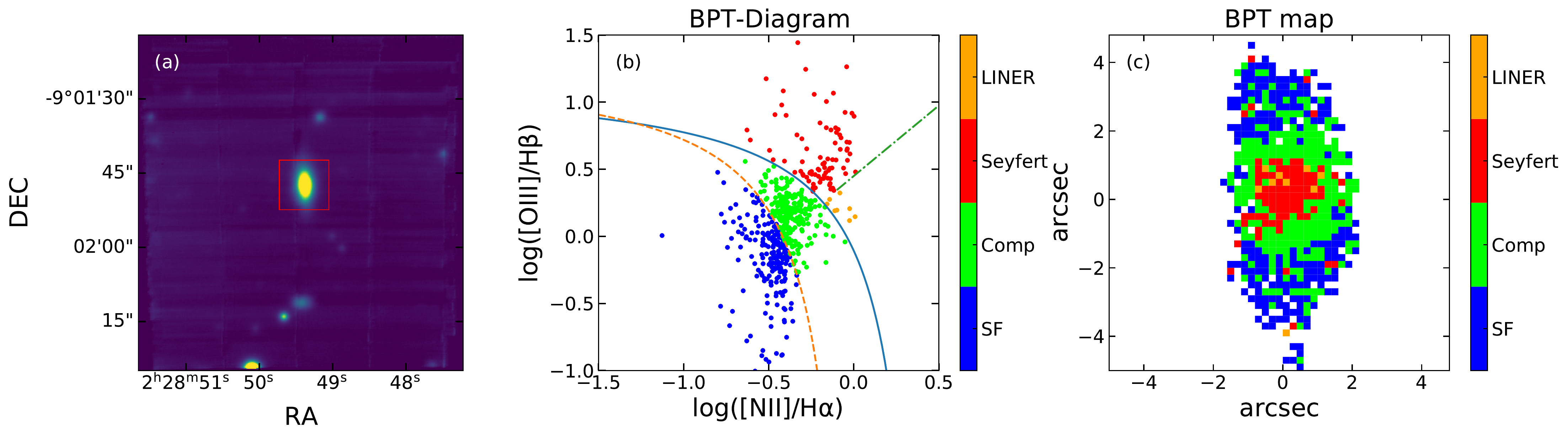}}
      \caption{White image and BPT map of SDSS J0228-0901. (a): The white image of SDSS J0228-0901, which is taken from MUSE observation. The FOV of MUSE is $1\arcmin \times 1\arcmin$, SDSS J0228-0901 is boxed by the red square with a size $10\arcsec \times 10\arcsec$. (b): BPT diagram. Here we use the $\rm log([\ion{O}{iii}]/H\beta) vs log([\ion{N}{ii}]/H\alpha)$ diagram to distinguish the type of each spaxel. The red pixels above the blue solid curve and the green dot-dash line represent the Seyfert region, the orange dots above the blur solid curve but below the green dot-dash line are the LINER regions, the blue pixels below the orange dashed curve mark the star-forming region, and the green pixels between the two curve is the composite region \citep{2001ApJ...556..121K,2003MNRAS.346.1055K,2007MNRAS.382.1415S}. We only classify the spaxels whose signal-to-noise ratio (SNR) of $\rm H\alpha$ emission line larger than 10 (i.e. $\rm SNR_{H\alpha > 10}$). (c): Two-dimensional BPT map. The color represents the same region as panel (b).}
         \label{fig:bpt map}
   \end{figure*}

The galaxy SDSS J0228-0901 is first identified as an AGN in \citet{2007ApJ...670...92G}. As listed in Table~\ref{tab:target_info}, its black-hole mass of $ (3.7\pm{0.3})\times10^5{\rm M_{\odot}}$ is measured through a single-epoch broad line width by \citet{2018ApJ...863....1C}. In order to search for and study outflows in this galaxy hosting such an intermediate-mass black hole, we carry out the MUSE IFU observations (PID: 0104.C-0181(C), PI: F. Bian). Figure~\ref{fig:bpt map} (a) shows the white image of our object. The red box in this panel marks the region that we used to do spectral analysis.

We use the well-known method to classify AGN region in the optical band: the BPT diagram \citep{1981PASP...93....5B,2001ApJ...556..121K,2003MNRAS.346.1055K, 2007MNRAS.382.1415S}. This diagram uses several emission line ratios in the optical band: log([\ion{N}{ii}]$\lambda$6583/H$\alpha$) vs. log([\ion{O}{iii}]$\lambda5007$/H$\beta$) to classify galaxies into Seyfert, LINER, star-forming and composite regions. Therefore, we classify each spaxel by the [\ion{N}{ii}]-BPT diagram in panel (b) and (c) of Figure~\ref{fig:bpt map}. As indicated by panel (c), the central regions of SDSS J0228-0901 are classified as Seyfert. As shown later in this study, the Seyfert regions overlap more or less with the outflow regions. 

\begin{table}
	\centering
	\caption{Basic properties of object SDSS J0228-0901}
	\label{tab:target_info}
	\begin{tabular}{lccr} 
		\hline
            \hline
               & Value & Reference\\
             \hline
		R.A.(J2000) & 02h28m49.5s & \citet{2020yCat.1350....0G}\\
		Decl.(J2000) & -09d01m53.8s & \citet{2020yCat.1350....0G}\\
		z & 0.07217 & \citet{2015ApJS..219...12A}\\
            Distance (Mpc) & 335.8 & This work\\
            $M_{\rm BH}/{\rm M_{\odot}}$ & $(3.67\pm0.27)\times10^{5}$ & \citet{2018ApJ...863....1C} \\
		$M_{\rm *}/{\rm M_{\odot}}$ & $(4.74\pm0.64)\times10^{9}$ & This work \\
            inclination angle ($^{\circ}$) & 59 & This work \\
		B/D & $0.26 \pm 0.02$ & This work \\
		Effective Radius (kpc) & $4.23\pm0.37 $  & This work\\
		${\rm log}(L_{\rm [\ion{O}{iii}]})\ {\rm (erg\;s^{-1})}$ & $40.68 \pm 0.12$ & This work\\
		$\rm \sigma_{H\alpha\;broad}\ {\rm (km\;s^{-1})}$ &  $451.23\pm 4.45$ & This work\\
		Eddington ratio ($\eta$) & 0.15 & This work \\
		\hline
	\end{tabular}
\end{table}

\subsection{MUSE Data Reduction}

MUSE (\citealp{2010SPIE.7735E..08B}) is one of the instruments onboard VLT. It has a spatial sampling of $\rm 0.2 \arcsec$ and a field of view of $\rm 1 \arcmin \times 1 \arcmin$ under the Wide Field Mode (WFM). This mode covers the wavelength range from $\rm 4650 \AA$ to $\rm 9300 \AA$, with a spectral resolution of about 2800. The observations with the WFM-NOAO-N mode were carried out on October 8, 20, and 21, 2019. The total on-source exposure time is 2400 sec. The typical seeing is around $\rm 1.4 \arcsec$. We adopt the standard \texttt{ESO Reflex} environment \citep{2013A&A...559A..96F} to produce the IFS datacube. Briefly, the science frames are first calibrated by bias, dark, and flat frames. Next, the arc frames are used for wavelength calibrations. After the sky subtractions, flux calibrations, and telluric corrections, the single science exposures are combined into the final datacube. The default settings of the pipeline are adopted during the data reduction processes.

\subsection{SDSS Data}

We also retrieve the broad band images from the Sloan Digital Sky Survey Data Release 15 (SDSS DR15, \citealp{2019ApJS..240...23A}). These images were taken on November 18, 2006, with a total exposure time of 53.9 secs and a seeing of $1.2 \arcsec$.  We further decompose the galaxy into bulge and disk components through the \texttt{GALFIT} software \citep{2010AJ....139.2097P}. The uncertainties of the photometry include the read noise and the sky background. 


\section{Data Analysis}

\subsection{Voronoi Binning}

To improve the signal-to-noise ratio (SNR) of our datacube, we rebin the spaxels by the \texttt{VorBin} software \citep{2003MNRAS.342..345C}. The datacube is binned according to the SNR of the H$\alpha$ emission line. To ensure sufficient SNR, we adopt a threshold of 10.

\subsection{Spectrum Fitting}

\begin{figure*}
   \resizebox{\hsize}{!}
            {\includegraphics[width=\textwidth]{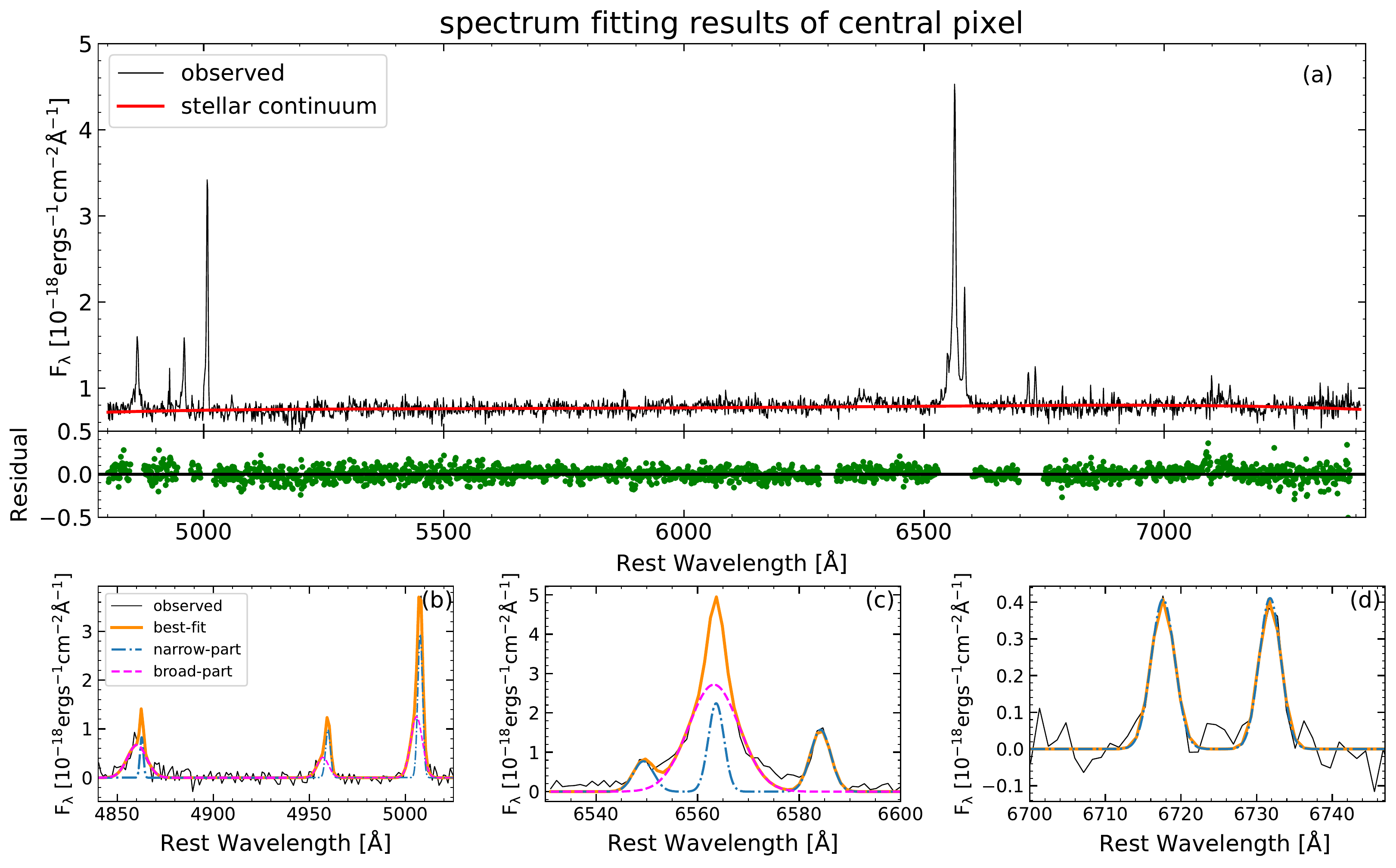}}
      \caption{Spectrum fitting results of the central spaxel. (a): The black line shows the observational data, the red line represents the stellar continuum fitted by \texttt{pPXF} software, and the green dots in the middle plane show the residuals.  (b)-(d) is the $\rm H\beta$ region, $\rm H\alpha$ region, [\ion{S}{ii}] doublet fitting results. Same as (a), the black line represents the observational data, the orange line is the best fit of the emission line, the blue dot-dashed line is the narrow component and the magenta dashed line is the broad component.}
         \label{fig:specfit}
   \end{figure*}

There is a strong blue-shifted wing in [\ion{O}{iii}]$\rm \lambda\lambda 4959,5007$ emission line as shown in Figure~\ref{fig:specfit}, indicating an ionized outflow in SDSS J0228-0901. Before measuring the emission line, we first employ the penalized Pixel-Fitting (\texttt{pPXF}; \citealp{2017MNRAS.466..798C}) software to fit the underlying stellar continuum. We adopt the single stellar population model MILES \citep{2010MNRAS.404.1639V} and a 4th-order polynomial. To align the spectral wavelength intervals with the model spectrum, we limit the wavelength range from 4800\AA\  to 7409\AA. 

For each emission line with a potential broad line component after subtracting the stellar continuum, we fit it with a single Gaussian model and multi-Gaussian model, respectively, through Python code \texttt{Scipy.curve\_fit}. If the former has a $\chi^2$ more than 20\% larger than the latter \citep{2022ApJ...937....7S}, we define that there is a broad component in this emission line. Through the above method, we fit a total of eight emission lines for our IFS datacube, including: $\rm H\alpha \lambda 6563$, $\rm H\beta \lambda 4861$, $\rm [\ion{N}{ii}] \lambda\lambda 6548,6583$, $\rm [\ion{O}{iii}] \lambda\lambda 4959,5007$, $\rm [\ion{S}{ii}] \lambda\lambda 6717,6731$. For the [\ion{O}{iii}]$\lambda 4959$ and [\ion{O}{iii}]$\lambda 5007$ doublet, we tied the line width and set the flux ratio of 1:3 \citep{2007MNRAS.374.1181D}.

As shown in Figure~\ref{fig:specfit}, the best fitting results are dominated by polynomials and do not show any strong absorption lines for the central spaxel, indicating the dominance of AGN power law emission over the stellar continuum emission at the center. As shown in Figure~\ref{fig:specfit} (b)-(d), there is a clear broad component in $\rm H\alpha$, $\rm H\beta$, and [\ion{O}{iii}] doublet.

\subsection{Dust Attenuation}

 We correct the dust attenuation based on the Balmer decrement and the attenuation curve from \citet{2001PASP..113.1449C}, with an assumption of $\rm A_V/E(B-V)=3.1$. For SDSS J0228-0901, we adopt the intrinsic ratio between the narrow components of $\rm H\alpha$ and $\rm H\beta$ to be 2.86 for star-forming regions and 3.1 for AGN regions \citep{2006MNRAS.372..961K}. 

\subsection{Outflow Velocity}

For the broad component in [\ion{O}{iii}] emission lines, we not only calculate its line center velocity (i.e. $V_{50}$) but also use the non-parametric approach from \citet{2014MNRAS.441.3306H} to evaluate the outflow velocity (i.e $V_{80}$). Where the $V_{80}$ represents at least 20\% of the flux of the emission line moving at this speed. In the non-parametric approach, we consider $W_{80}$ as the width which contains $80\%$ of the total flux of the broad component of the emission line. And then we use the equation from \citet{2019ApJ...884...54M} to define:
\begin{equation} 
    V_{80} = -V_0 + \frac{W_{80}}{2},
\end{equation}
where $V_0$ is the line center velocity offset between narrow and broad components, and for a single Gaussian profile: $W_{80}=1.09{\rm FWHM}$. 

\subsection{Rotation Curve}

\begin{figure}
   \centering
   \includegraphics[width=\columnwidth]{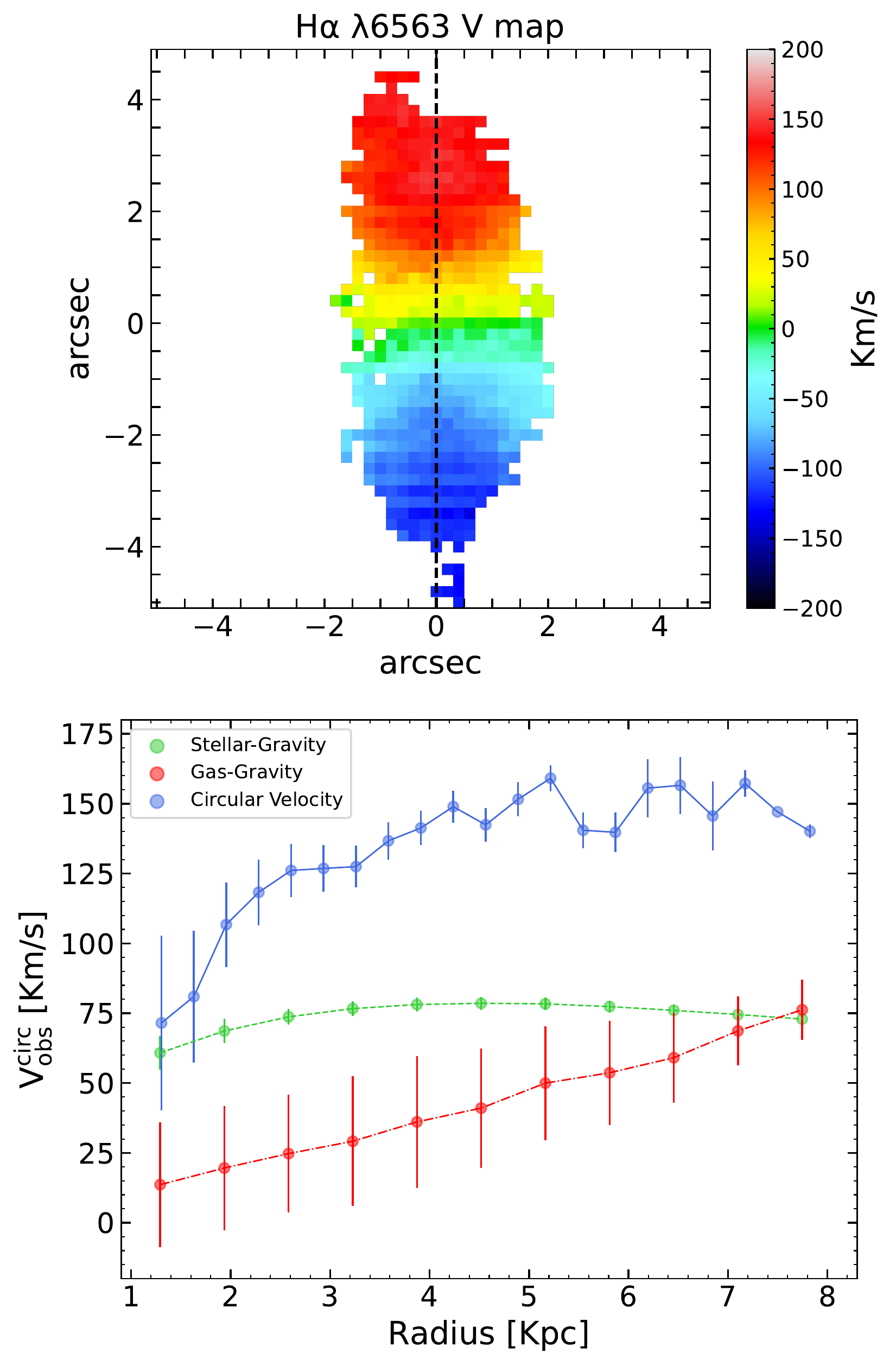}
      \caption{(a): Line center velocity map of $\rm H\alpha$ narrow components, the black dashed line along the major axis. Setting the $\rm H\alpha$ emission line SNR threshold to 10. (b): The contribution of different components to the total rotation curve. The blue line shows the circular velocity profile which is from the rotation curve from the $\rm H\alpha$ velocity with the correction of asymmetric drift. The green dotted line represents the contribution of the stellar component, while the red dot-dashed line of the gas component.}
         \label{fig:vel-rc}
   \end{figure}
   
\begin{figure*}
   \resizebox{\hsize}{!}
             {\includegraphics[width=\textwidth]{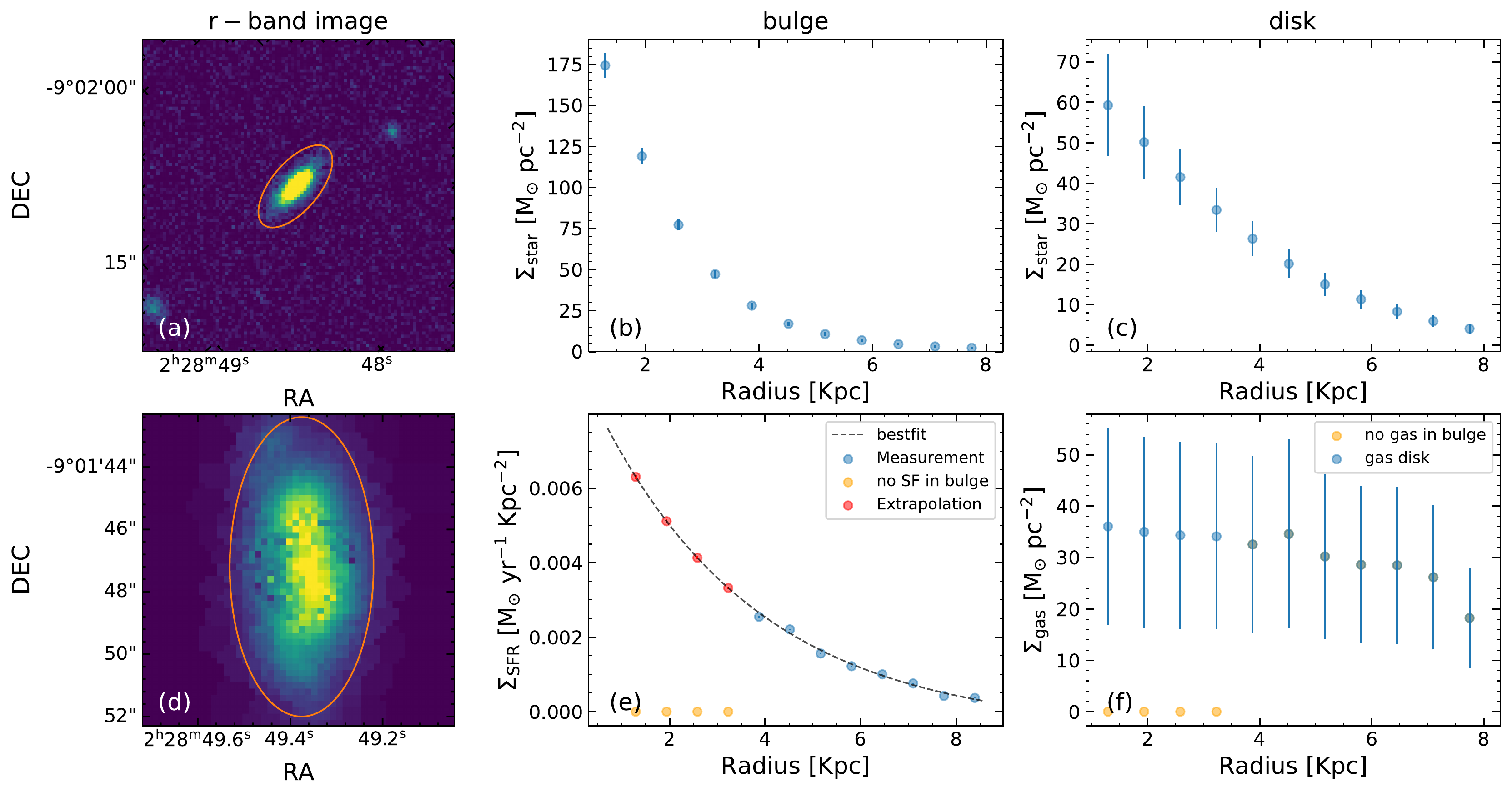}}
      \caption{Optical $r$-band image and $\rm H \alpha$ integrate flux map of SDSS J0228-0901. (a), (d): The $r$-band image from SDSS DR15 and the $\rm H \alpha$ integrate flux map from MUSE observation. The orange ellipse represents the outer bounds of aperture photometry. (b), (c): Radial profile of stellar mass surface density of galaxy bulge and disk components. (e): Radial profile of star formation rate, the sky-blue dots are the star formation rate surface density that we measure from the $\rm H \alpha$ luminosity, the black dashed line is the best-fit exponential model, and the red points are obtained by extrapolation from the exponential model. The orange dots are presented as the model that there is no star formation in the inner regions. (f): Radial profile of total gas mass surface density derive from the extended Schmidt law. The sky-blue dots mark the exponential star formation model, and the orange dots show the model that no star formation in the inner regions.}
         \label{fig:photometry}
   \end{figure*}

To evaluate if the outflow can escape the galaxy gravitational potential well, we measure the total rotation curve from the line center velocity map of the H$\alpha$ narrow component. Here the H$\alpha$ velocity map includes the narrow line region (NLR) of AGN. We estimate the inclination angle by measuring the ellipticity through fitting the surface brightness image with the python package \texttt{photutils} \citep{larry_bradley_2022_6825092}. Figure~\ref{fig:vel-rc} (a) shows the velocity map, and we measure the total rotation curve of the galaxy along its major axis which is the black dashed line in this figure. 

The observed rotation curve contains contributions from the gravity of stars, gas, and dark matter. We measure the stellar contribution from the stellar mass surface density profile that is based on the SDSS DR15 $g$ and $r$ band images through the aperture photometry. As the bulge and disk have different 3-D structure and thus different gravitational potential well for a given surface density. We first adopt \texttt{GALFIT} \citep{2010AJ....139.2097P} to decompose the image data by fitting $g$ and $r$ images with a bulge of S$\rm \acute{e}$rsic index 2 and an exponential disk model while estimating the error as described in \S~2.2. Afterward, we use the python package \texttt{photutils} to perform aperture photometry on two decomposed components, while setting the outer boundary of the photometry around 8 kpc. We then estimate the stellar mass profile from the surface brightness profile of the $g, r$ band with the equation from \citet{2018ApJ...853..149S} that are based on the SDSS galaxies:
\begin{equation}
    {\rm log} M_{*} + M_{\rm r-band}/2.5 = 1.13 + 1.49\times(g-r),
\end{equation}
where $M_{\rm r-band}$ is the $r$-band absolute magnitude in AB, and $g-r$ is color in AB. Figure~\ref{fig:photometry} (a)-(c) shows the $r$-band image, the stellar mass profile of bulge and disk, respectively, where the orange ellipse in panel (a) marks the outer boundary and reaches about 8 kpc. To measure the stellar contribution to the rotation curve, we use the \texttt{ROTMOD} task of \texttt{GIPSY} \citep{1983MNRAS.203..735C} by assuming a "SECH-SQUARED" vertical distribution with a scale height of 0.2 kpc. The error of the stellar rotation curve is dominated by the uncertainties of stellar mass distribution, and this effect is simulated through the Monte-Carlo method. As a result, the green dashed curve in Figure~\ref{fig:vel-rc} (b) is the stellar rotation curve, its inner most radius reaches about 1.2 kpc as limited by the spatial resolution of the SDSS image.

It is out of the current capability to obtain both spatially-resolved \ion{H}{i} gas and molecular gas kinematics for this dwarf galaxy with a luminosity distance of about 300 Mpc. As a result, we estimate the total gas surface density through the star formation law, especially the extended Schmidt law that shows a tight relationship between star formation rate (SFR), gas, and stellar mass down to very low gas densities \citep{2011ApJ...733...87S, 2018ApJ...853..149S, 2022arXiv221107215D}. 
To estimate the SFR surface density, we first draw the flux map of $\rm H\alpha$ emission line narrow component. To exclude the effect of AGN, we mask the regions which are classified as AGN in the BPT map. Figure~\ref{fig:photometry} (d) shows the H$\alpha$ narrow component flux map, where the orange ellipse represents the outer boundary. We then estimate the SFR through the equation from \citet{1998ARA&A..36..189K}:
\begin{equation}
    {\rm SFR} ({\rm M_{\odot}yr^{-1}}) = 7.9 \times 10^{-42}L(\rm H \alpha) ({\rm ergs\; s^{-1}}),
\end{equation}
where $L({\rm H \alpha})$ is the flux of the $\rm H\alpha$ narrow component in ${\rm ergs\;s^{-1}}$.For the inner region, we assume two models to evaluate the SFR in the bulge: (1) no star formation in the bulge regions. (2) we adopt a gas disk model so that the SFR surface density follows an  exponential model from the inner regions to the outer regions. Figure~\ref{fig:photometry} (e) shows the SFR surface density evaluated from the two models above, where the error is mainly from the flux error of the $\rm H\alpha$ emission line. With the derived SFR and stellar mass distribution, we measure the gas mass distribution from the equation from \citet{2018ApJ...853..149S}:
\begin{equation}
    \rm \Sigma_{SFR} = 10^{-4.76}(\Sigma_{star}^{0.5}\Sigma_{gas})^{1.09},
\end{equation}
where the $\rm \Sigma_{SFR}$ is the SFR surface density in
$\rm M_{\odot}\;yr^{-1}\;kpc^{-2}$, $\rm \Sigma_{star}$ and $\rm \Sigma_{gas}$ are the stellar and gas mass surface densities in $\rm M_{\odot}\;pc^{-2}$, respectively. As a result, the gas distribution is shown in Figure~\ref{fig:photometry} (f), where the error contains the error from SFR, stellar mass profiles and the intrinsic scatter of around 0.23 dex for the extended Schmidt law. We then use the same method to measure the gas contribution to the rotation curve as we measure the stellar rotation curve. The most significant difference between the gas rotation curve measured by the two models is that the rotation velocity of the first method is much slower than the gas disk model in the inner bulge regions. The overall gas contribution to the circular velocity from the two models is minimal, so we ensure that the dark matter halo (DMH) rotation curve also becomes flat when the total rotation curve is flattened.
 
To evaluate the escape velocity of the gravitational potential of the galaxy, we also need to measure the circular velocity. However, as the gas collision, part of the gravitational potential of the galaxy is consumed by the random motion of gas referred to as the asymmetric drift effect \citep{2015AJ....149..180O}. We follow the method of \citet{2021ApJ...909...20S} to correct the effect of asymmetric drift. For a hydrostatic equilibrium disk, the circular velocity satisfies:
\begin{equation}
     V_{\rm circ}^2 = V_{\rm rot}^2 + V_{\rm P}^2,
\end{equation}
where $ V_{\rm rot}$ is the rotation velocity from observations and $ V_{\rm P}$ is the velocity related to the gas random motion as driven by pressure. The $ V_{\rm P}$ is related to the gas density and gas dispersion:
\begin{equation}
     V_{\rm P}^2 = -R\sigma_{\rm v}^2 \frac{\partial {\rm ln}(\rho \sigma_{\rm v}^2)}{\partial R}.
\end{equation}
If adding another assumption that the scale height ($z_0$) of the gas disk is a constant, i.e. $ \mathrm{d}{\rm ln}(z_0)/\mathrm{d}R = 0$, then the equation will become:
\begin{equation}
    V_{\rm P}^2 = -R\sigma_{\rm v}^2 \frac{\partial {\rm ln} (\sigma_{\rm v}^2{\Sigma_{\rm obs}} {\rm cos}i)}{\partial R},
\end{equation}
where $\rm \Sigma_{obs}$ is the gas surface density from the observation, $\rm \sigma_v$ is the gas velocity dispersion as obtained from the narrow component of the H$\alpha$ emission line, and $i$ is the inclination angle. The final rotation curve after correcting the pressure support is shown in Figure~\ref{fig:vel-rc} (b). Recently, \citet{2023MNRAS.tmp..857D} investigated the relationship between the measured rotation curve and circular velocity in dwarf galaxies by simulations. They pointed out that the environmental effects can cause significant differences between the two, which may not be corrected fully by the asymmetric drift effect. However, the optical image of SDSS J0228-0901 shows that it is an isolated dwarf galaxy that lacks obvious merging or interacting features. On the other hand, we are mainly interested in the flat part of the circular velocity that infers the escape velocity. In their sample, the rotation curve and circular velocity do show a better agreement in the flat part. Therefore, these environmental effects should not affect our results significantly.


\section{Results and Discussion}


\begin{table*}
 \centering
 \caption{Comparison of physical properties between SDSS J0228-0901 and other samples from literature.} 
 \label{tab:different samples}
 \resizebox{1.0\textwidth}{!}{
 \begin{tabular}{lccccccc}
  \hline
  \hline
  Samples & ${\rm log} M_{*}/M_{\rm \odot}$ & ${\rm log} M_{\rm BH}/M_{\rm \odot}$ & ${\rm log } L_{\rm AGN}\ (\rm erg\;s^{-1})$ & Eddington Ratio ($\eta$) & $V_{\rm 80}\;({\rm km\;s^{-1})}$ & Effective Radius (kpc) & $\rm log SFR$ ($\rm M_{\odot}\;yr^{-1}$)\\
   \hline
   SDSS J0228-0901 & 9.6 & 5.5 & 42.8 & 0.15 & 471 & 4.2 & -0.48 \\
   \citet{2020ApJ...905..166L} &  8.77 - 9.97 & 4.8 - 5.8 & 42.0 - 43.9 & 5e-2 - 3.2 & 300 - 800 & 0.6 - 2.0 & -2.0 - -0.22 \\
   \citet{2015MNRAS.446.2186M} &  10.5 - 11.5 & 7.4 - 8.9 & 44.4 - 45.5 & 7e-3 - 0.39 & 369 - 962 & -- & 1.56 - 2.73\\ 
   \citet{2018ApJ...864..124K} &  -- & 6.2 - 9.3 & 42.9 - 45.3 & 7e-4 - 0.79 & -- & 2.3 - 13.2 & --\\         
        \hline 
 \end{tabular}}
\end{table*}

\subsection{Spatial Distributions and Kinematics of Outflow}

\begin{figure*}
   \resizebox{\hsize}{!}
             {\includegraphics[width=\textwidth]{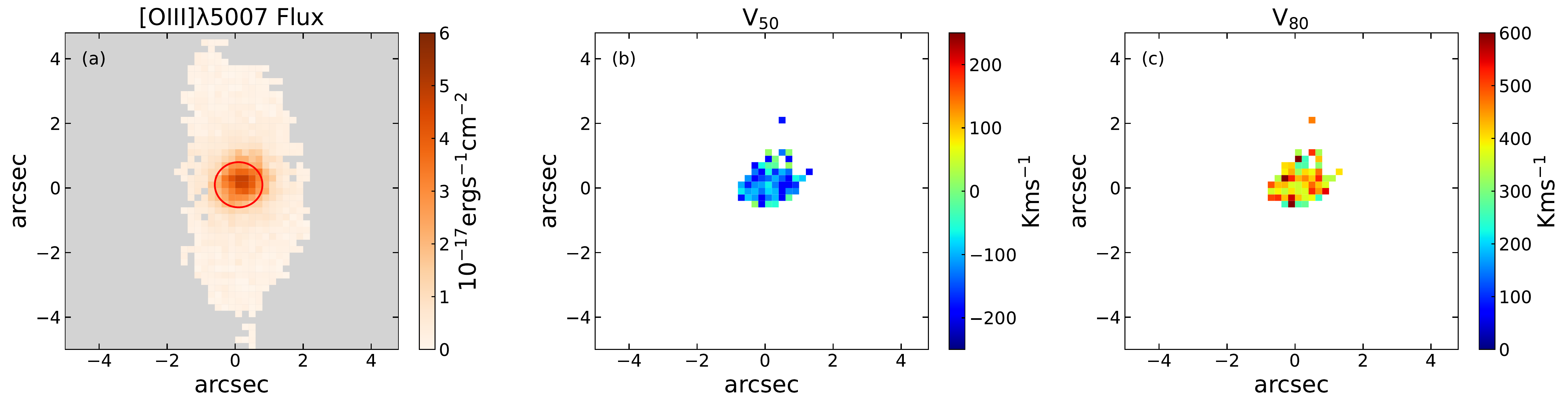}}
      \caption{The spatially resolved physical properties of galaxy SDSS J0228-0901.  (a): The $\rm [\ion{O}{iii}]$ total flux map with the correction of dust attenuation, we use the same SNR cut as described before. Due to the weak emission and seeing smearing effect, we can not reach the threshold ($\rm 10^{-16}\; erg\;s^{-1}\;cm^{-2}\;arcsec^{-2}$) of the extended narrow line region (ENLR) size measured in \citet{2019MNRAS.489..855C}. Hence we plot half of the FWHM of PSF as the red circle in panel (a) to represent the upper limit of the ENLR size. (b) The line center velocity ($ V_{\rm 50}$) of the broad component of $\rm [\ion{O}{iii}]$ emission line, and the blue color represent the blue shifted velocity, the maximum $V_{\rm 50}$ reach to around $\rm -200\ km\;s^{-1}$. (c): The outflow velocity ($V_{\rm 80}$) map defined as $V_{\rm 80}$ which is calculated by the same method in \citet{2019ApJ...884...54M}, the maximum $V_{\rm 80}$ along minor axis reach to $\rm 471\ km\;s^{-1}$. Both the $V_{\rm 50}$ and $V_{\rm 80}$ do not decrease with the radius, and the outflow can extend to galactic scale with a size of around 3 kpc by rough estimation. }
         \label{fig:spatial-distribution}
   \end{figure*}

\begin{figure}
   \centering
   \includegraphics[width=\columnwidth]{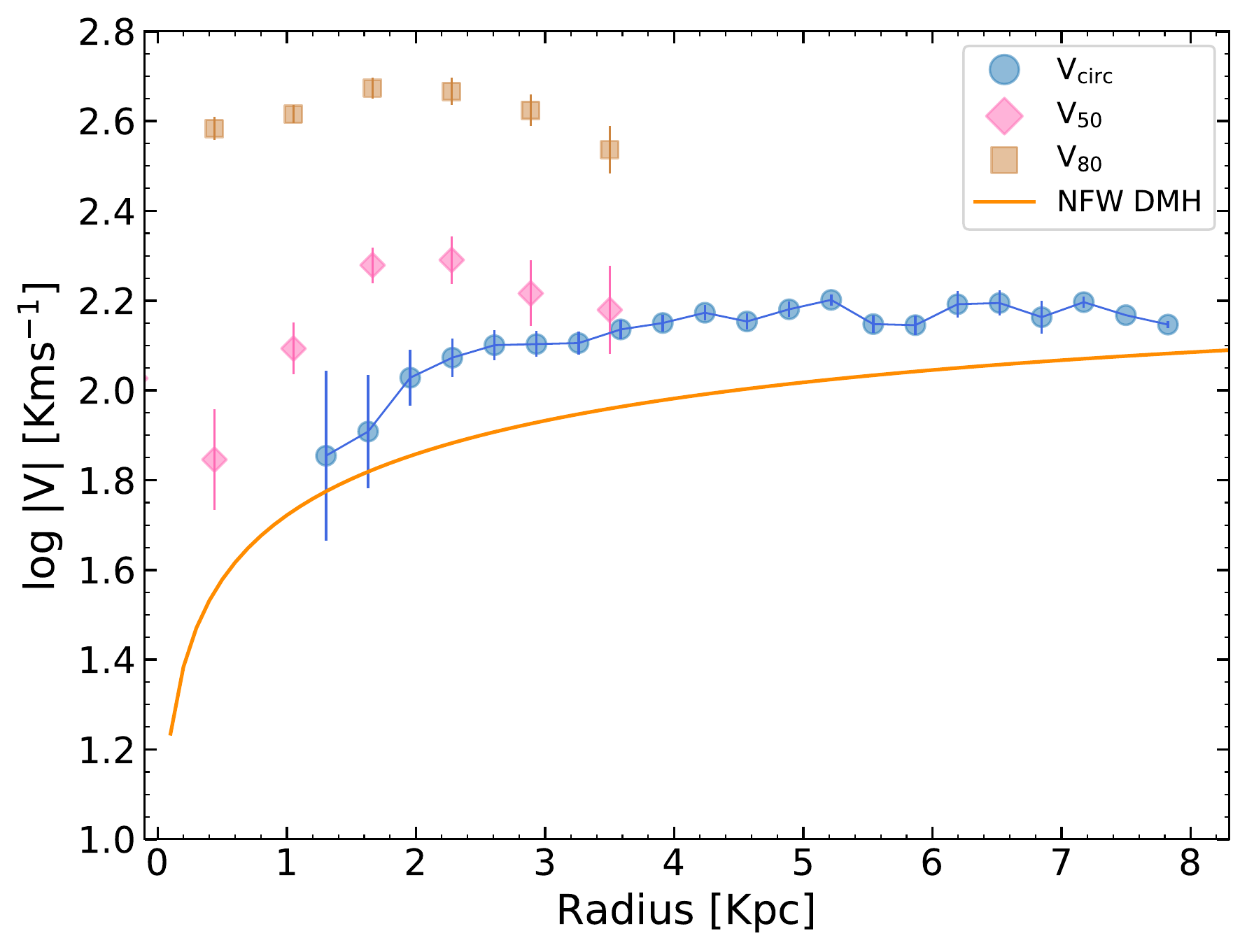}
      \caption{The spatially extended outflow velocity vs total rotation curve in log space. The blue curve is the total rotation curve, the pink diamonds mark the line center velocity ($V_{\rm 50}$) while the brown squares are the outflow velocity ($V_{\rm 80}$). The dark orange curve represents the circular velocity profile of an NFW DMH.}
         \label{fig:outflow-rc}
   \end{figure}

\begin{figure*}
  \resizebox{\hsize}{!}
            {\includegraphics[width=\textwidth]{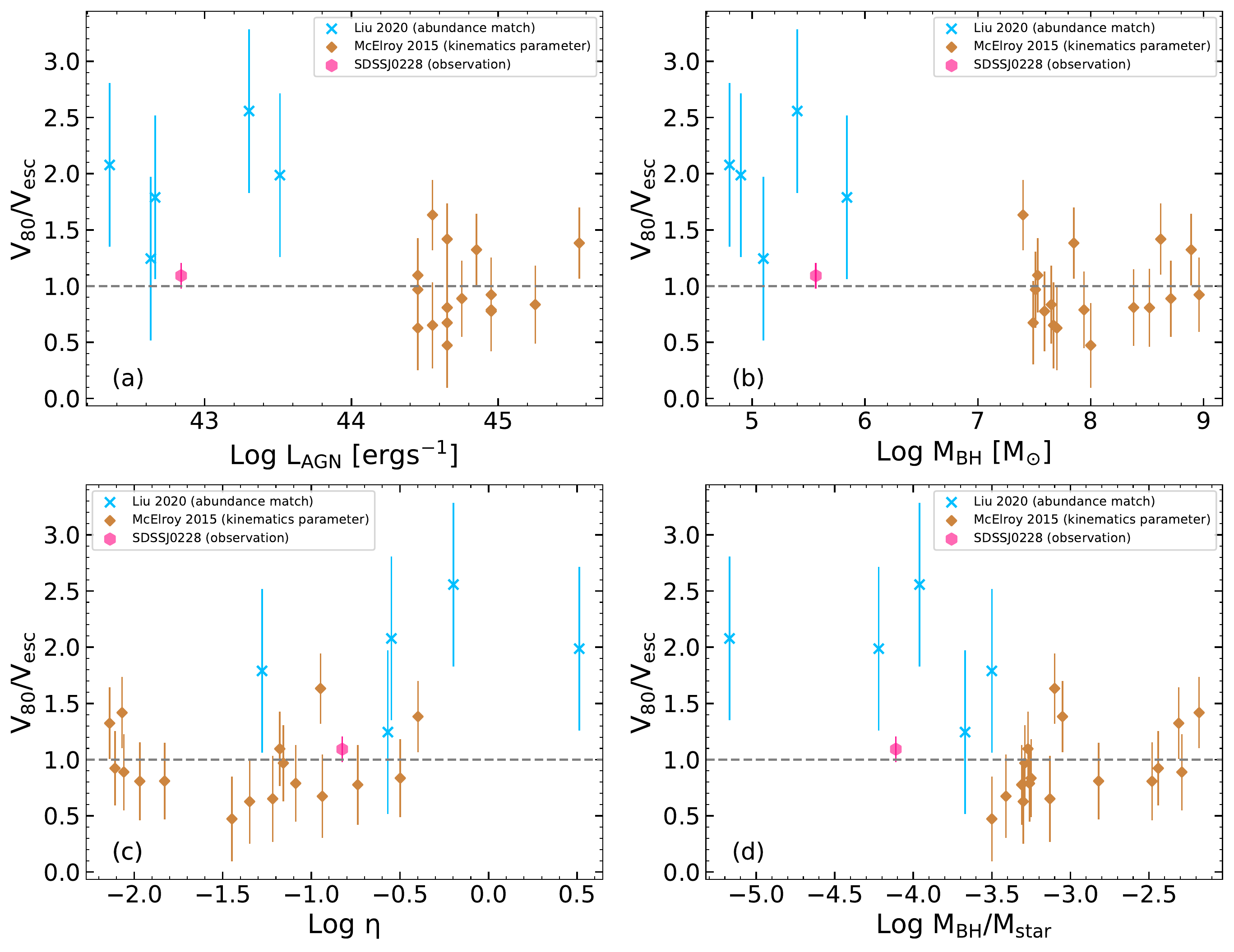}}
      \caption{Outflow capability vs. AGN properties. We utilize the ratio between the maximum outflow velocity and the escape velocity ($V_{\rm 80}/V_{\rm esc}$) to describe the escape capability of outflow. (a): $V_{\rm 80}/V_{\rm esc}$ as a function of AGN luminosity, (b): $V_{\rm 80}/V_{\rm esc}$ as a function of SMBH mass. (c): $V_{\rm 80}/V_{\rm esc}$ as a function of Eddington ratios. (d): $V_{\rm 80}/V_{\rm esc}$ as a function of the ratio between SMBH mass and stellar mass}. In each subplot, the pink hexagon represents the galaxy SDSS J0228-0901, the sky-blue cross is the dwarf galaxies from \citet{2019ApJ...884...54M} and \citet{2020ApJ...905..166L}, and the 
      brown square mark the massive galaxies from \citet{2015MNRAS.446.2186M}. The gray dashed line represents where $V_{\rm 80}/V_{\rm esc} = 1$.
         \label{fig:AGN-relation}
  \end{figure*}

We estimate the kinematic properties of the ionized gas outflow as shown in Figure~\ref{fig:spatial-distribution} (b) and (c). The outflow regions are almost all blue-shifted and their velocities ($V_{\rm 50}\ {\rm and}\ V_{\rm 80}$) do not decrease with the radius. Figure~\ref{fig:outflow-rc} compares the radial profiles of two outflow velocities with the rotation curve. The $V_{\rm 80}$ remains almost constant with radius, and it is much larger than the circular velocity.

We also use the ratio of outflow velocity and the halo escape velocity ($V_{\rm 80}/V_{\rm esc}$) to evaluate if the outflow can escape from the gravitational potential of the host galaxy. We select a total of 23 sources that include both dwarf galaxies and massive galaxies for comparison. Five dwarf galaxies with AGN are from \citet{2019ApJ...884...54M} and \citet{2020ApJ...905..166L}. They have available SMBH mass, $V80$, and halo escape velocity measurements. Massive galaxies with AGN are from \citet{2015MNRAS.446.2186M}. We list a series of physical properties of SDSS J0228-0901 and comparsion samples in Table~\ref{tab:different samples}.

We estimate the escape velocity of SDSS J0228-0901 through two methods. (1) By assuming the density profile of both baryonic and dark matter with an isothermal sphere model which truncates at a radius $r_{\rm max}$, the escape velocity is related to $V_{\rm circ}$ and $r_{\rm max}/r$ as described in \citet{2020A&ARv..28....2V}:
\begin{equation}
    V_{\rm esc}(r) = V_{\rm circ}\sqrt{2\left [1+ {\rm ln}(\frac{r_{\rm max}}{r})\right ]}.
\end{equation}
For different $r_{\rm max}/r$, the $V_{\rm esc}/V_{\rm circ}$ can vary from 2.6 to 3.3 with different $r_{\rm max}/r$. We use the mean factor of 3 to estimate the escape velocity to be about $\rm 427.2\ km\;s^{-1}$, i.e. $V_{\rm esc} = 3V_{\rm circ}$. (2) The escape velocity can also be derived from the gravitational potential directly:
\begin{equation}
    V_{\rm esc}(r)^{2} = 2\times |\Phi(r)|.
\end{equation}
Here we assume the DMH as a Navarro-Frenk-White (NFW) halo model \citep{1997ApJ...490..493N}, and estimate the halo mass from maximum circular velocity \citep{2011ApJ...740..102K}. As a result, the estimated halo mass ($M_{200}$) is around $10^{11.95}\ {\rm M_{\odot}}$ and the escape velocity is around $\rm 435\  km\;s^{-1}$ at $R = R_{\rm out}$ ($R_{\rm out}$: the radius of the outflow region.). Finally, we take the average of the two methods as the result, i.e. $V_{\rm esc} = (431.4 \pm 17.9)\ {\rm km\;s^{-1}}$. We notice that both methods produce a consistent escape velocity, indicating that the NFW DMH model is reasonable to describe the dark matter distribution in SDSS J0228-0901. In Figure~\ref{fig:outflow-rc}, the dark orange curve represents the circular velocity profile of the NFW DMH. We calculate this profile through the derived NFW circular profile\citep{2021ApJ...909...20S}:
\begin{equation}
    V_{\rm NFW} (R) = V_{200} \sqrt{\frac{{\rm ln}(1 + cx) - cx/(1+cx)}{x[{\rm ln} (1+c) - c/(1+c)]}}),
\end{equation}
where $V_{200}$ is the velocity at $R_{200}$, i.e. $V_{200} = R_{200}h$, c represent the concentration with $c = R_{200}/R_s$, and $x = R/R_{200}$. The halo concentration parameter ($c$) is estimated from its halo mass ($M_{200}$) following the approximation for distinct halo \citep{2011ApJ...740..102K}:

\begin{equation}
    c(M_{200}) = 9.60 (\frac{M_{200}}{10^{12}h^{-1}{\rm M_{\odot}}}).
\end{equation}

As for the subsample of massive galaxies, the circular velocity is determined by a kinematic parameter $S_{0.5}$ that is the combination of the regular rotation ($ V_{\rm rot}$) and random motion ($\sigma_{\rm gas}$). To measure the kinematic parameter $S_{0.5}$, we adopt a tight correlation between $S_{0.5}$ and stellar mass ($M_*$) \citep{2007ApJ...660L..35K}: 
\begin{equation}
    {\rm log} (M_*/{\rm M_{\odot}}) = 1.93 {\rm log} S_{0.5} + 0.30.
\end{equation}
Then, according to the stellar mass from \citet{2015MNRAS.446.2186M}, the circular velocity is estimated as $V_{\rm circ} = \sqrt{2}S_{0.5}$. In the subsample of dwarf galaxies, \citet{2019ApJ...884...54M} used the abundance matching method \citep{2013MNRAS.428.3121M} to estimate the mass of DMH and assume an NFW dark matter density profile \citep{2001MNRAS.321..155L} to derive the escape velocity. 

The escape velocity error of SDSS J0228-0901 is mainly from the error of the circular velocity profile which is described in \S~3.5. For the escape velocity estimated from the kinematic parameter, the scatter of the relation between $S_{0.5}$ and $M_*$ dominates the error of the velocity. For dwarf galaxies, assuming a standard deviation of 0.2 dex for the abundance-match based ${\rm log}M_{200}$ \citep{2022arXiv221102665X}, the uncertainty of $V_{\rm circ}$ is derived from the relation between $V_{\rm circ, max}$ and $M_{200}$ \citep{2011ApJ...740..102K}, which gives a fractional error of  $\Delta V_{\rm circ} / V_{\rm circ} \approx 0.7$. Errors of outflow velocity and $W80$ in the total 23 sources are from the spectral fitting.

Figure~\ref{fig:AGN-relation} (a), (b) show $V_{\rm 80}/V_{\rm esc}$ as a function of the AGN luminosities and SMBH masses. The $V_{\rm 80}/V_{\rm esc}$ of SDSS J0228-0901 is $1.09\pm0.04$, larger than most massive galaxies. These results suggest that the AGN-driven outflow in dwarf galaxies is able to escape from the gravitational potential of the dark matter halo. This result is consistent with \citet{2020ApJ...905..166L} although their escape velocity
measurements have much larger errors. Figure~\ref{fig:AGN-relation} (c) shows $V_{\rm 80}/V_{\rm esc}$ as a function of the Eddington ratios. Although there is a significant selection bias in dwarf galaxies toward higher Eddington ratios as compared to massive galaxies, no clear trend is seen that the AGN-driven outflow escapes more easily in AGNs with high Eddington ratios. Figure~\ref{fig:AGN-relation} (d) shows $V_{\rm 80}/V_{\rm esc}$ as a function of the ratio between the SMBH mass and stellar mass of hosts. There is no clear trend suggesting that the AGN-driven outflows escape more easily in more massive SMBHs at a given stellar mass.

 Besides Figure~\ref{fig:AGN-relation}, we also compare other properties among these samples in Table~\ref{tab:different samples}. For the environments, we further check the optical image in SDSS. Dwarf galaxies are all isolated and lack merging or interacting features. While about half of massive galaxies show clear merging or interacting features, galaxies with interacting pairs or tidal tails do not show significantly higher outflow velocity \citep{2015MNRAS.446.2186M}. On the other hand, these merging or tidal features may affect the estimation of stellar mass and escape velocity. For the SFR, \citet{2020ApJ...905..166L} estimated the SFR upper limit from the [\ion{O}{ii}] luminosity, which indicated that their dwarf galaxies span a large range from green-valley galaxies to starburst galaxies. \citet{2015MNRAS.446.2186M} provided the total H$\alpha$ luminosity data, from which we convert to the SFR following the method in \citet{1998ARA&A..36..189K}, and find that these massive galaxies have slightly higher SFRs consistent with the SF main sequence and starburst galaxies.

Other studies of outflows in larger samples also show $V_{\rm out}/V_{\rm esc}$ as a function of the stellar mass, indicating that ionized gas outflows in dwarf galaxies are more likely to escape from their hosts \citep{2014A&A...568A..14A, 2019MNRAS.486..344R, 2019MNRAS.490.4368S, 2022ApJ...929..134X}. While this conclusion is supported by current observations, it is important to note that AGNs in dwarf galaxies are selected to have high Eddington ratios. It is unclear whether outflow can be launched to a similar velocity when the accretion rate is low in dwarf galaxies.

\subsection{Outflow size vs. AGN luminosity }

\begin{figure}
   \centering
   \includegraphics[width=\columnwidth]{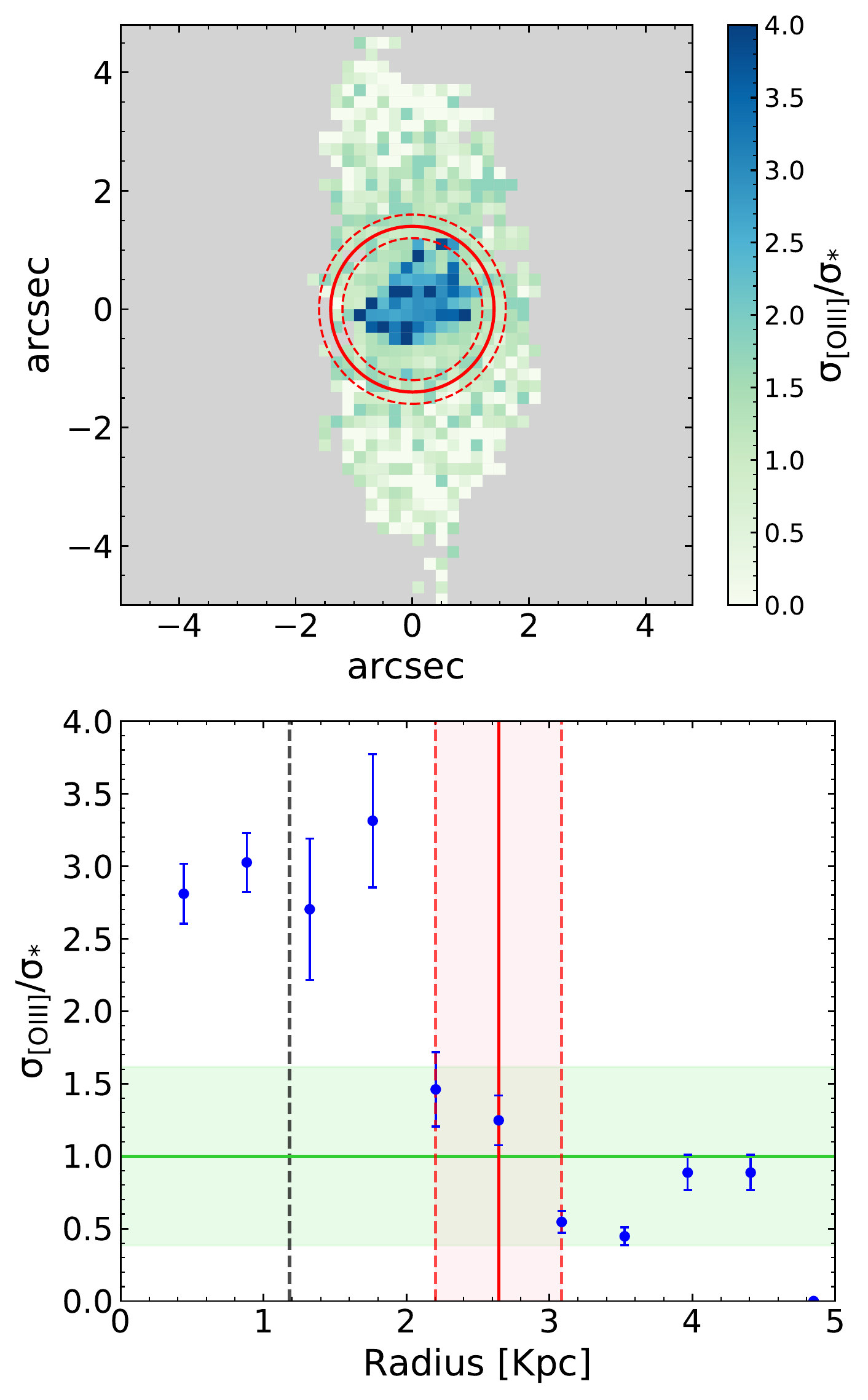}
      \caption{Upper: Spatial distribution of [\ion{O}{iii}] velocity dispersion which is normalized by the stellar velocity dispersion, the ionized gas outflow size is indicated by the red circle with $1\sigma$ error. Bottom: the normalized [\ion{O}{iii}] velocity dispersion profile along the major axis, here we adopt the same cut as the rotation curve, i.e. SNR of $\rm H\alpha$ larger than 10. The green line and light green region represent the stellar dispersion with $1\sigma$ error, the vertical red line is the outer boundary of the outflow region, the two vertical dashed red lines and the pink area show the $1\sigma$ error, while the vertical black dashed line represents the PSF size (FWHM/2). After subtracting the seeing smearing, we obtain the final outflow size: $ R_{\rm out} = (2.45\pm0.44)$ kpc.}
         \label{fig:outflow-size}
   \end{figure}
   
\begin{figure}
    \centering
    \includegraphics[width=\columnwidth]{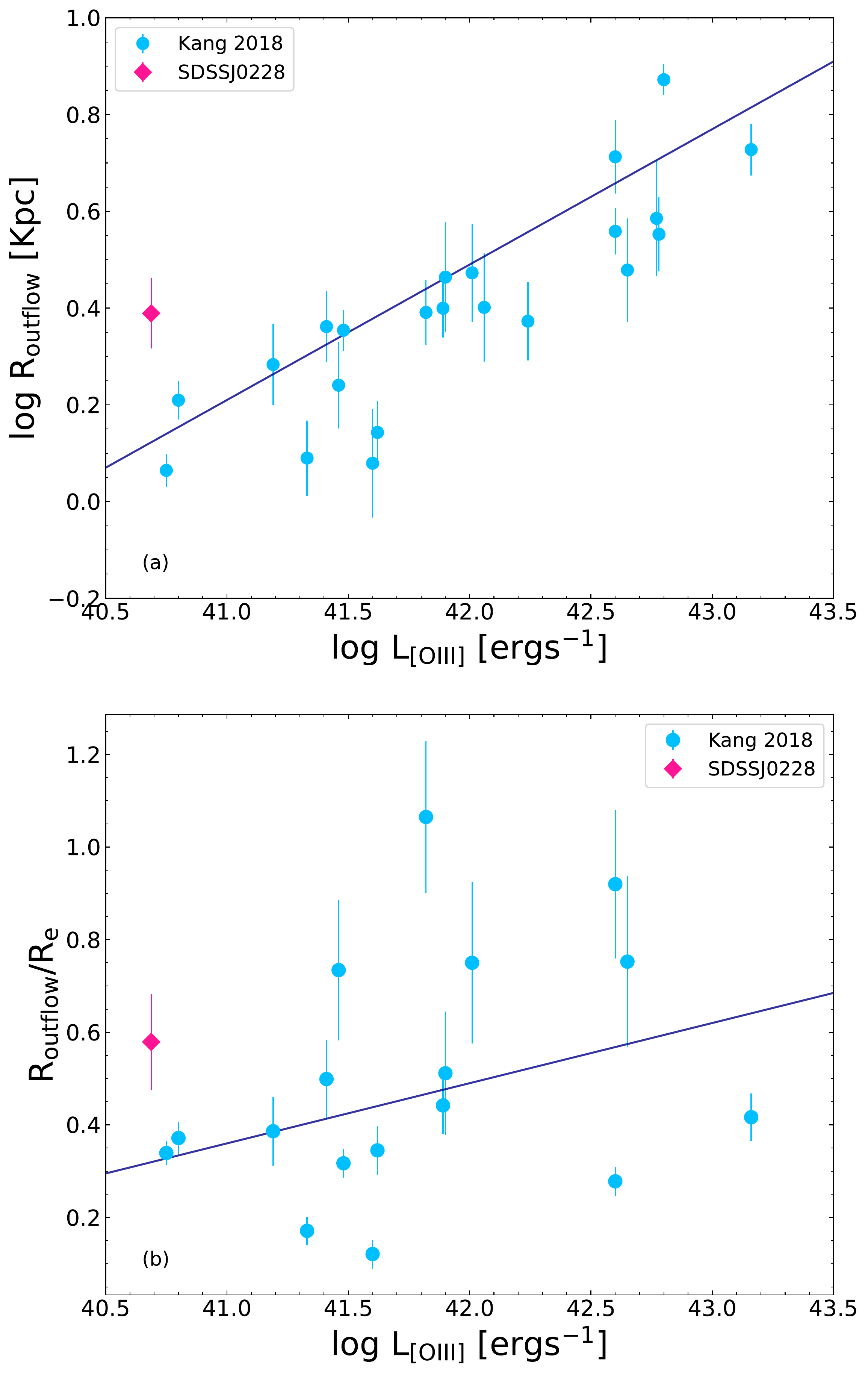}
    \caption{Upper: Outflow size vs [\ion{O}{iii}] luminosity, both the blue dots and the best-fit relation with the dark blue line are from \citet{2018ApJ...864..124K}, the pink marker is galaxy SDSS J0228-0901. The outflow size of SDSS J0228-0901 is significantly larger than predicted by the relation. Bottom: Normalized outflow size by effective radius ($ R_{\rm out}/R_{\rm e}$) vs [\ion{O}{iii}] luminosity, the symbols are same as the upper panel. The dark blue line marks the best-fit relation of the sky-blue data points.}
    \label{fig:Kang18}
\end{figure}

In addition to the outflow velocity, the size of the outflow region is another indicator to reflect how the AGN exerts feedback on its host galaxy. Numerous works have studied outflow sizes using spatially resolved data, revealing that outflow sizes range from the sub-kpc \citep{2018PASJ...70...93K, 2019MNRAS.482.3915B} to several kpc \citep{2014MNRAS.441.3306H, 2015MNRAS.446.2186M, 2018ApJ...864..124K}. Two types of methods have been used to measure outflow sizes. One is to define the outflow region based on the line fluxes such as signal-to-noise ratio or surface brightness thresholds. The second is based on the kinematics of emission lines, e.g. the outflow region
is identified as the presence of a broad line component whose $\rm \sigma_{emline} > 250\; km\; s^{-1}$, or as the velocity dispersion of the emission line equal to the stellar velocity dispersion, i.e. $\rm \sigma_{emline}=\sigma_*$ \citep{2018ApJ...864..124K}.

Previous works have compared the outflow size using different methods, and pointed out that the flux based method may significantly over-estimate the outflow size \citep{2018ApJ...856..102F, 2018ApJ...864..124K}. Therefore we estimate the outflow size of SDSS J0228-0901 by using the kinematic edge in \citet{2018ApJ...864..124K}. They studied 23 luminous ($L_{\rm \ion{O}{iii}}>5\times10^{41}\;{\rm erg\;s^{-1}}$) type 2 AGN that host relatively massive SMBH in the local universe ($M_{\rm BH} \sim 10^{6.2-9.3}\; {\rm M_{\odot}}$) and defined the ionized gas outflow region as where the velocity dispersion of  $\rm [\ion{O}{iii}] \lambda5007$ emission line is larger than the stellar velocity dispersion. The derived outflow sizes of their samples cover from 0.60 to 7.45 kpc. We use the same method to estimate the outflow size for SDSS J0228-0901. However, since we do not detect any absorption line in the spectrum of SDSS J0228-0901, it is impossible to obtain the stellar velocity dispersion directly. Fortunately, previous works proposed to use the flux weighted velocity dispersion of the narrow component of the emission line as the proxy of stellar velocity dispersion, such as H$\alpha$ or [\ion{O}{iii}] \citep{2016ApJ...817..108W,2017ApJ...845..131K,2018ApJ...864..124K}. Here, we substitute the stellar velocity dispersion with the flux weighted velocity dispersion of the narrow component of [\ion{O}{iii}]$\lambda$5007, and convert $\sigma_{\rm [\ion{O}{iii}]narrow}$ to $\sigma_{*}$ by dividing with a factor of $(1.09 \pm 0.32)$, to give $\sigma_{*}$ of around $\rm 63\ kms^{-1}$. We note that this correction factor has a slight effect on the outflow size measurement but does not change our results. The AGN luminosity are converted from [\ion{O}{iii}] luminosity \citep{2009A&A...504...73L}

The upper panel of Figure~\ref{fig:outflow-size} shows the two-dimensional spatial distribution of $\rm [\ion{O}{iii}] \lambda5007$ velocity dispersion distribution that is normalized with the stellar velocity dispersion, while the bottom panel shows the normalized $\rm [\ion{O}{iii}] \lambda5007$ velocity dispersion along the major axis.We estimate the outflow size of $R_{\rm out} = 2.45\pm0.44$ kpc by subtracting the PSF size (black dashed line). \citet{2018ApJ...864..124K} obtains a good relation between the outflow size and [\ion{O}{iii}] luminosity:
\begin{equation}
    {\rm log}(R_{\rm out}) = (0.28\pm0.03)\ {\rm log} L_{[\ion{O}{iii}]} - (11.27\pm1.46).
\end{equation}
Figure~\ref{fig:Kang18} (a) clearly shows that the kinematic outflow size of SDSS J0228-0901 locates above the relationship.

Furthermore, to compare the outflow size among galaxies with different sizes, we normalize the outflow size by the effective radius ($R_{\rm e}$) of galaxies. Here we use \texttt{photutils} to measure the surface brightness profile from the SDSS $r$ band image by aperture photometry. Then the surface brightness profile is fitted by an S$\rm \acute{e}$rsic model \citep{1968adga.book.....S} which contains $R_{\rm e}$:

\begin{equation}
    I(R) = I_{\rm e} {\exp} \left\{ -b_{\rm n}	\left[ \left(\frac{R}{R_{\rm e}}\right)^{1/n} -1 \right]\right\}.
\end{equation}
During the fitting, we free three parameters: $I_{\rm e}$, $R_{\rm e}$, and S$\rm \acute{e}$rsic index $n$, where $b_{\rm n}$ is tied with $n$ as $b_{\rm n} \approx 1.9992n - 0.3271$ \citep{2019PASA...36...35G}. Given that the average FWHM of PSF is $\rm 1.26 \arcsec$, if the effective radius is smaller than the PSF size, we then use the PSF size as the upper limit of the effective radius. Figure~\ref{fig:Kang18} (b) shows the result where we exclude sources classified as quasars in archive data. Finally, we plot 17 galaxies together with SDSS J0228-0901. It seems the normalized outflow size of SDSS J0228-0901 is not so stand out from the other galaxies but is slightly larger than other AGN in the low [\ion{O}{iii}] luminosity regime. The above two results indicate that the mechanical feedback from the central AGN is more substantial in SDSS J0228-0901.

\begin{figure}
    \centering
    \includegraphics[width=\columnwidth]{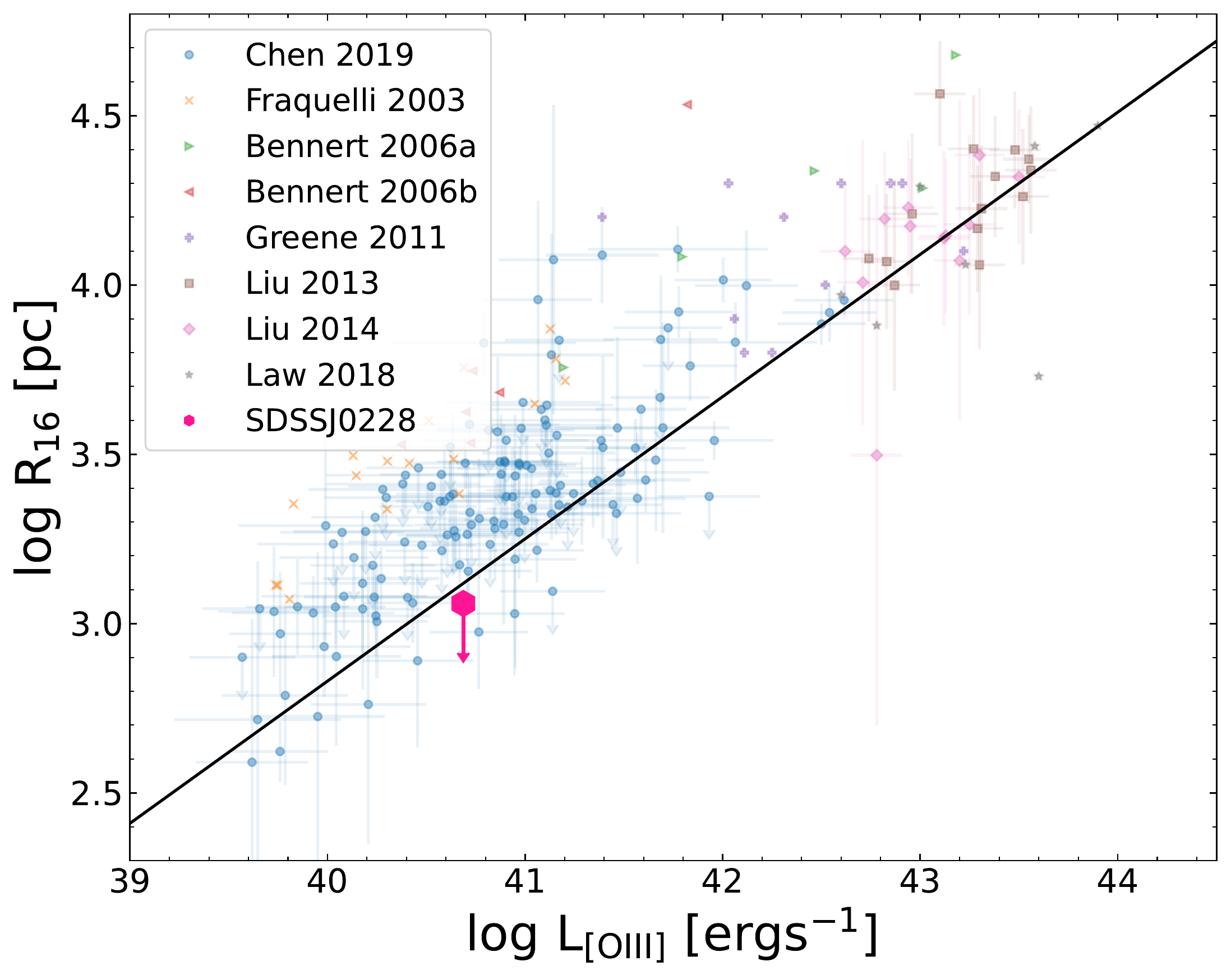}
    \caption{Extended Narrow Line Region size vs [\ion{O}{iii}] luminosity. All the background dots are same as the Figure 6. in \citet{2019MNRAS.489..855C}, the best-fit relation with a slope of $(0.42\pm0.02)$. The pink hexagon represents the upper limit ENLR size of SDSS J0228-0901, we set the upper limit as half of the FWHM of median PSF as $\rm 1.1\ kpc$.}
    \label{fig:Chen19}
\end{figure}

Since the AGN feedback is not only through the mechanical process (i.e. jets and outflows) but also via the radiative process i.e. photoionization regions as excited by AGN \citep{2015ARA&A..53..115K}. The size of the extended narrow line region (ENLR) is an indicator of the radiative effect. Similar to the size measurements of outflow regions, there are also different methods to determine the size of ENLR. For the Seyfert galaxies, we follow the method in \citet{2019MNRAS.489..855C}, and use the same threshold of $\rm 10^{-16}\ erg\;s^{-1}\;cm^{-2}\;arcsec^{-2}$. However, Figure~\ref{fig:spatial-distribution} (a) indicates that the surface brightness across the whole galaxy is all below this threshold, indicating a tiny size if adopting the above threshold. In Figure~\ref{fig:Chen19}, we adopt half of the PSF FWHM as the upper limit of the ENLR size and found that its upper limit does not deviate significantly from the relation.

Owing to the difference in gas motion, there is a large difference between outflow size and ENLR size. From observations, the outflow size is always smaller than the size of ENLR \citep{2018ApJ...856..102F, 2018ApJ...864..124K}. This indicates that the mechanical feedback is weaker than the radiative feedback in most Seyfert galaxies. However, for SDSS J0228-0901 we find a larger outflow size than the ENLR size, which indicates the mechanical feedback is stronger than the radiative feedback in this object.
 
\subsection{Gas phase metallicity of star-forming region and AGN NLR}

\begin{figure*}
   \resizebox{\hsize}{!}
             {\includegraphics[width=\textwidth]{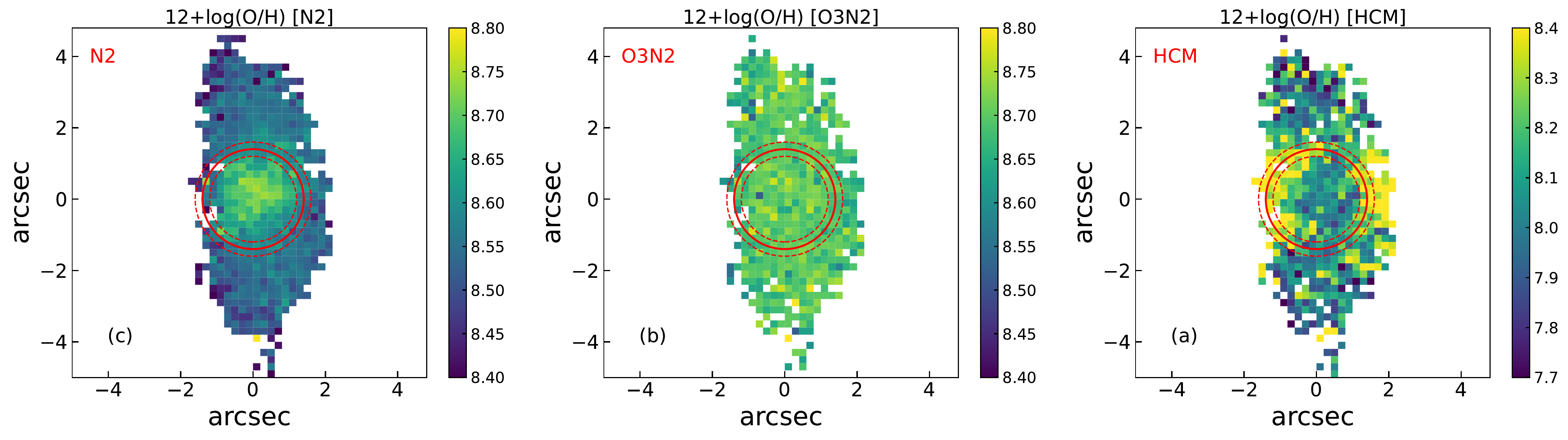}}
      \caption{Two-dimensional gas-phase metallicity map of SDSS J0228-0901. We estimate the gas-phase metallicity through three methods. (a) The N2 calibrator proposed by \citet{2020MNRAS.492.5675C}. (b) The O3N2 calibrator proposed by \citet{2019MNRAS.485..367K}. (c) The python script \texttt{HII-CHI-MISTRY} (\texttt{HCM}) from the photoionized model \citep{2014MNRAS.441.2663P}. In each panel, the red circle represents the size of outflow regions with $\rm 1 \sigma$ error.}
         \label{fig:mental}
   \end{figure*}

Metallicity  is an important property of ionized gas in galaxies. However, the metallicity of gas that is ionized by the AGN radiation is difficult to measure. This is because most of the emission line diagnostics are usually adapted for star-forming regions. Therefore we estimate the gas-phase metallicity through 3 different methods. (1) N2 calibrator which is used in Seyfert 2 AGNs and proposed by \citep{2020MNRAS.492.5675C}:
\begin{equation}
    12+ {\rm log}\left (\frac{\rm O}{\rm H} \right ) = 8.69 + {\rm log} \left (\frac{Z}{Z_{\odot}} \right),
\end{equation}
where $Z/Z_{\odot} = 4.01^{N2}-0.07$, and $N2 = {\rm log}([\ion{N}{ii}]\lambda6584/{\rm H\alpha})$. This calibrator is also adopted to AGN NLRs and LINERs for $0.3<(Z/Z_{\odot})<2.0$. (2) O3N2 calibrator which is used in diffused ionized gas and LINERs as proposed by \citep{2019MNRAS.485..367K}:
\begin{equation}
    12+ {\rm log}\left (\frac{\rm O}{\rm H} \right ) = 7.673+0.22 \times \sqrt{25.25 - 9.072 \times O3N2} + 0.127 \times O3,
\end{equation}
where $O3N2 = {\rm log}([\ion{O}{iii}]\lambda5007/{\rm H\beta} \times {\rm H\alpha}/[\ion{N}{ii}\lambda6584])$, and $O3 = {\rm log}([\ion{O}{iii}]\lambda5007/{\rm H\beta})$. (3) \texttt{HII-CHI-MISTRY} (\texttt{HCM}) software that uses multiple strong collisionally exciting lines in the optical band and combines them with photo-ionized models to calculate the gas-phase metallicity \citep{2014MNRAS.441.2663P}. We adopt the \texttt{HCM} software to obtain the spatially resolved gas-phase metallicity of SDSS J0228-0901 from the inner AGN regions to outer star-forming regions. Within the wavelength coverage of MUSE, we have five emission lines: $\rm [\ion{O}{iii}] \lambda\lambda 4959, 5007$, $\rm [\ion{N}{ii}] \lambda 6584$, $\rm [\ion{S}{ii}] \lambda\lambda 6715,6731$, and [\ion{S}{ii}] lines for which we sum them together as $\rm [\ion{S}{ii}] \lambda 6725$. The fluxes of all these emission lines are normalized by the flux of the narrow component of $\rm H\beta$ emission line. To adapt the photoionized model better, we exclude the broad component from the $\rm H\alpha$ and $\rm H\beta$ emission line and the blue-shifted wing of [\ion{O}{iii}] doublet. 

Figure~\ref{fig:mental} shows the metallicity distribution estimated by the three methods above. SDSS J0228-0901 shows an overall flat radial metallicity distribution, a general property seen in dwarf galaxies. Even in AGN NLRs, the metallicity is not apparently elevated. However, there is a significant difference between the two optical line calibrators and the \texttt{HCM} software. As shown in Figure~\ref{fig:mental} (a), (b), both N2 and O3N2 show weak or no negative gradients along the major and minor axis and do not present metal enrichment around outflow edges on the minor axis direction. But in Figure~\ref{fig:mental} (c), along the minor axis, regions that are outside the galaxy disk but within the outflow regions have enhanced metallicity. In these regions, gas may be dominated by outflowing gas that is more metal rich. The \texttt{HCM} software is also adopted to various galaxies, including the star-forming galaxies, Seyfert 2 AGNs, and LINERs \citep{2021MNRAS.505.4289P}. In that work, their Seyfert 2 AGN and LINER sub-samples do not show a clear mass metallicity relation (MZR), and the metallicity in their star-forming galaxies also presents a significant deviation in the low mass end in MZR. As for SDSS J0228-0901, a dwarf galaxy with a Seyfert 1 AGN, the metallicity inferred from the \texttt{HCM} software also shows a large deviation in MZR (according to the MZR, the metallicity is about 8.7 as stellar mass is around 9.6). As a result, we prefer  results from the two emission line calibrators, suggesting that AGN does not launch metal rich outflowing gas in our object.

%

 
 \subsection{Electron Density of star-forming region and AGN NLR}
 
 \begin{figure}
    \centering
    \includegraphics[width=\columnwidth]{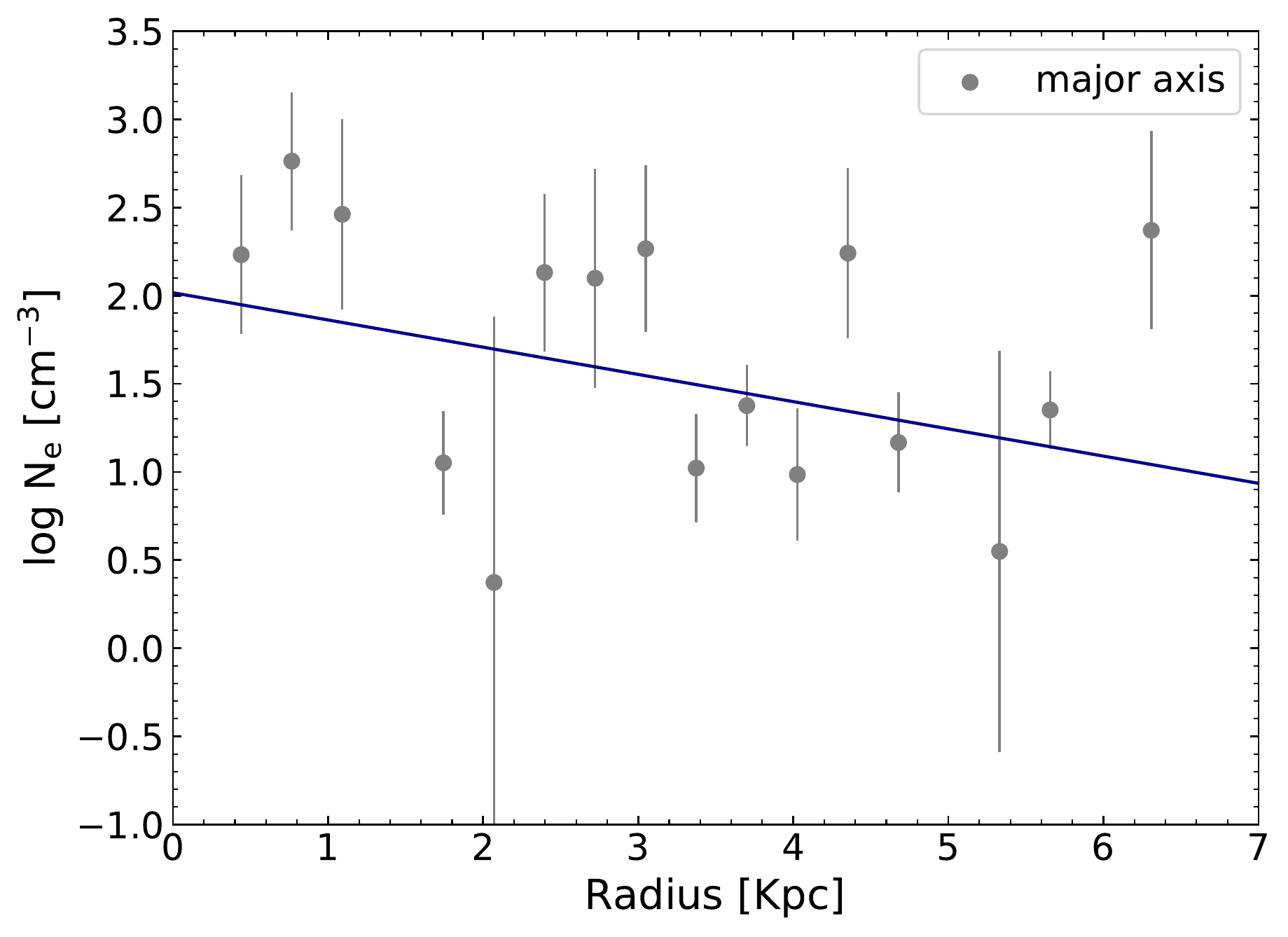}
    \caption{The electron density along the major axis. The electron density is estimated by the ratio of $\rm [\ion{S}{ii}]$ doublet. While the dark blue solid line is the best fit line of the electron density gradient with a slope of $\rm -0.15 \pm{0.08}$. }
    \label{fig:ne}
\end{figure}

The electron density ($n_{\rm e}$) of ionized gas can be derived from the ratio between $\rm[\ion{O}{ii}] \lambda\lambda 3726, 3729$ or $\rm [\ion{S}{ii}] \lambda\lambda 6717,6731$ doublets \citep{2016ApJ...816...23S}, which covers a $n_{\rm e}$ range of $\rm 50-5000\ cm^{-3}$. Here we estimate the electron density of NLR through the flux ratio of $\rm [\ion{S}{ii}]$ doublet. As shown in Figure~\ref{fig:ne}, the $n_{\rm e}$ along the major axis exhibits a shallow negative gradient with the best fit slope of $(-0.15 \pm{0.08})$. 

Some previous studies about the AGN-driven ionized gas outflow assume $n_{\rm e}$ as a constant in the outflow regions \citep{2014MNRAS.441.3306H,2015MNRAS.446.2186M,2016A&A...592A.148K}, but the electron density is very different between the NLR and the outflow regions. While other works with IFS data confirm that the gas in the outflow regions is denser than in the NLR, and the electron density in NLR also decreases with radius \citep{2014A&A...568A..14A,2018A&A...618A...6K,2020MNRAS.494.5396B}. Our result gives a conclusion that the electron density in NLR decreases with radius too. However, due to the lack of the broad component of $\rm [\ion{S}{ii}]$ emission line, we can not obtain the electron density of the outflow regions. The result also shows that the $n_{\rm e}$ of NLR is not much different from normal galaxies which indicates the outflow may not affect the electron density of NLR distribution of the host galaxy.


\section{Conclusions}

With VLT-MUSE observations, we discover a fast AGN-driven outflow in a dwarf galaxy SDSS J0228-0901 that hosts an intermediate-mass black hole (${M_{\rm BH}\sim 10^5\;{\rm M_{\odot}}}$). By comparing with other galaxies that host AGN-driven outflows, we find that the outflow in dwarf galaxies can escape when their central black-holes accrete at high rates. And here are our main conclusions.

(1) In the BPT diagram and BPT map, the detected outflow regions largely overlap with the AGN NLR regions, which supports that the outflow is driven by the central AGN.

(2) The outflow in SDSS J0228-0901 is blue-shifted, with the outflow velocity ($V_{80}$) of $\rm 471\; km\;s^{-1}$, while the escape velocity is around $\rm 431\; km\;s^{-1}$. This indicates that the outflow will escape from the gravitational potential of the galaxy and dark matter halo.

(3) To further investigate how AGN feedback affects its host galaxy in our case. We estimate the outflow size and extended narrow line region (ENLR) size, and find that the outflow size is larger than the ENLR size, indicating the mechanical feedback process is stronger than the radiative feedback process. 

(4) From the two-dimensional distribution of gas-phase metallicity, we do not find the metallicity enhancement in the outflow regions.

(5) Due to the lack of the broad component in $\rm [\ion{S}{ii}] \lambda\lambda 6717,6731$, we cannot obtain the electron density of the outflow regions. But we do calculate the electron density of the NLR by using the narrow component of $\rm [\ion{S}{ii}]$ doublet and find a negative gradient along the major axis with the best-fit slope of $\rm -0.15\pm0.08$.

\section*{Acknowledgements}

We thank the referee for helpful and constructive comments. Z.Z. and Y.S. acknowledges  the support from the National Key R\&D  Program of
China No. 2022YFF0503401, the National Natural Science Foundation of
China (NSFC  grants 11825302, 12141301, 12121003). This work has been supported by the New Cornerstone Science Foundation through the XPLORER PRIZE. Based on observations made with ESO Telescopes at the La Silla Paranal Observatory under programme ID 0104.C-0181(C).

\section*{Data Availability}

The data underlying this article will be shared on reasonable request to the corresponding author.



\bibliographystyle{mnras}
\bibliography{dAGN_outflow} 








\bsp	
\label{lastpage}
\end{document}